# Free Energy Rate Density and Self-organization in Complex Systems

Georgi Yordanov Georgiev[1,2,3*], Erin Gombos[1,4], Timothy Bates[1], Kaitlin Henry[1], Alexander Casey[1,5], Michael Daly[1,6],

[1]Physics Department, Assumption College, 500 Salisbury St, Worcester, MA, 01609, USA
[2]Physics Department, Tufts University, 4 Colby St, Medford, MA, 02155, USA
[3]Department of Physics, Worcester Polytechnic Institute, Worcester, MA, 01609, USA
[4]Current address: National Cancer Institute, NIH, 10 Center Drive, Bethesda, MD 20814
[5]Current address: University of Notre Dame, Notre Dame, IN 46556, USA
[6]Current address: Meditech, 550 Cochituate Rd, Framingham, MA 01701, USA
[*]Corresponding author. Emails: ggeorgie@assumption.edu; georgi@alumni.tufts.edu

**Abstract.** One of the most important tasks in science is to understand the self-organization's arrow of time. To attempt this we utilize the connection between self-organization and non-equilibrium thermodynamics. Eric Chaisson calculated an exponential increase of Free Energy Rate Density (FERD) in Cosmic Evolution, from the Big Bang until now, paralleling the increase of systems' structure. We term these studies "Devology". We connect FERD to the principle of least action for complex systems, driving their increase of action efficiency. We study CPUs as a specific system in which the organization, the total amount of action and FERD are connected in a positive feedback loop, providing exponential growth of all three and power law relations between them. This is a deep connection, reaching to the first principles of physics: the least action principle and the second law of thermodynamics. We propose size-density and complexity-density rules in addition to the established size-complexity one.

## 1 Introduction

After many years of study, the processes of self-organization of complex systems still do not have a satisfactory description. Non-equilibrium thermodynamics is essential to understand self-organization [1-3]. The driver towards higher levels of structure and organization in complex systems and in Cosmic Evolution has been recognized as the density (time and mass) of free energy flowing through a system [4-6]. Energy differences (gradients) create forces which move flows of matter, doing work to minimize constraints to motion, thus carving the flow channels along which the product of time and energy per event is minimized [7-10]. Other aspects of the principle of least action driving self-organization have been considered [13-16]. In our research program we set to investigate the entire chain of self-organizing events in systems spanning from the atoms to the society [7-10]. This is the goal of many scientists working in the fields of complexity, cosmic evolution, big history and others. This paper is one of the steps in this process answering the question whether there are other characteristics than level of organization and size of a system that are useful in describing the process of self-organization. Is free energy rate density, used by Chaisson, correlated to action efficiency as a measure of organization? Is the growth of free energy rate density exponential in time and does it fit in the positive feedback model between level of organization and size of a complex system? Will we be able to use those three measures as tools to characterize complex systems when partial information is available, and deduce them from each other?

Our approach is to study the minimization of average physical action per unit motion (action efficiency, quality) and the maximization of the total action for all motions in complex systems (quantity) [7-10]. We defined organization as action efficiency ($\alpha$) and size of a system as the total amount of action in it (Q) and used the principle of least action as a driving force for increase of $\alpha$ [7-10]. We also posited a maximum action principle, where Q in a self-organizing complex system tends to a maximum, i.e. for the unit action to decrease, the total action has to increase in order to minimize the constraints to motion further. We have shown that the quality ($\alpha$) depends on quantity (Q) and vice versa (the size-complexity rule) and when one is increased or decreased the other is affected in the same way, i.e. they are in a positive feedback loop [10]. This dependence is a major driving force and a mechanism of progressive development measured as the increase of action efficiency in complex systems.

Previously we described how in complex systems, elements cannot move along their least possible action paths that characterize their motion outside of systems, because of obstacles to the motion (constraints) [7-10]. We used variational approaches to optimization in complex systems, which are the least and most action principles mentioned above [7-10]. We extended the principle of least action as: complex systems are attracted toward a state with least average action per one motion given those constraints [7-10]. This is congruent with the Hertz's principle that objects move along paths with least curvature [11] and the Gauss principle that they move along the paths of least constraint [12]. We extended these principles for complex systems that the elements do work on the constraints to minimize them, reducing the curvature and the amount of action spent for unit motion. The tendency to move along geodesics, drives the flows of elements is a system to remove constraints from their paths in order to achieve the state of smallest possible product of time and energy. This is what we term self-organization. For this work the elements need energy, and the higher the free energy rate density, the more the work can be done, therefore the faster the self-organization. The new geodesics of the elements in the curved by the constraints to motion space are the paths with minimum action. The paths of least constraint are the flow paths in the system [7-10]. Therefore we defined organization (the action efficiency of the complex system) as the state of the constraints to motion determining the average action per one element of the system and one of its motions [10]. We posited a flow network representation of a complex system, where the flows are necessary to equilibrate any energy differences, attracted by their final state - that of thermodynamic equilibrium. In our model, the flow is of events, but not of energy or matter. Each element in a complex system is the smallest mobile unit in the system and usually moves in a flow channel along a network of paths (edges) between the starting and ending points (nodes) which are sources and sinks in the flow network. In CPUs, one unit of motion (event) is a single computation in which electrons flow from the start node to the end node [10].

The increase of FERD in complex systems is allowed by the increase of α driven by the principle of least action and of Q, driven by the principle of most action. In this paper, we understand the processes of progressive increase of level of organization, as a connected system of physics laws, which when put together yield complex systems that we observe around us. If any of those three principles is taken separately: the Least Action Principle (LAP), the Most Action Principle (MAP) and the principle of increase of Free Energy Rate Density (FERD), they do not lead to a self-organizing complex system. Only when they are connected in the same system in a positive feedback loop, acting together, they yield the amazing diversity of complex systems that we see in the world around us. We can use a new term, Devology ( *dev-* from "development", *-evo-* from "evolution" and *-logia* "study of") for a study of development of organization in complex systems in Cosmic Evolution, from the Big Bang to Humankind [4].

## 2  Model

Previously we connected in a positive feedback loop organization (α) and size (Q) of a complex system, leading to an exponential increase of both and to a power law relationship between the two, which matched well with data for CPUs [10]. We proposed that this feedback loop is the major mechanism of accelerated rate of self-organization and evolution of complex systems [10].

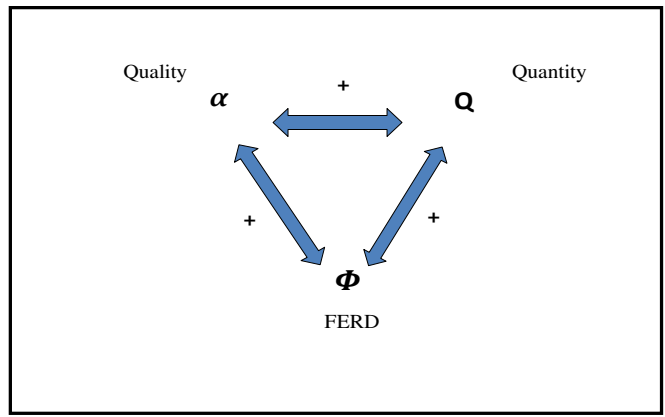

**Fig. 1.** A positive feedback model between α, Q and **Φ**. This loop can be described with a system of ordinary differential equations, as in [10], which solve to an exponential growth of each of the three and power law dependences between each two of them.

Here we show how non-equilibrium thermodynamics connects to this model, through measurement of what Eric Chaisson terms Free Energy Rate Density (**Φ**), as a distance from thermodynamic equilibrium, which he uses as a measure for Cosmic Evolution [4]. On Fig. 1 we include **Φ** in the model of positive feedback between α and Q developed earlier [10]. In this expanded model, all three are in a positive feedback, which as shown in [10] leads to exponential solutions of the differential equations for the involved quantities for each. When the exponential equations are combined, they yield a power law relation between them. Therefore for this paper it is enough to demonstrate with data whether the relationship of **Φ** to either α or Q is a power law in order to connect all three in a positive feedback loop with the mentioned outcomes.

## 3 Data and Methods

CPUs are organized flow systems, where the events are well defined as computations and precise data for time and energy per event available over the entire period of their existence. CPUs are a good model system, because they are analogous to all other complex systems: they perform events, consume energy and increase their time and energy efficiency over time – evolve. Therefore they are an excellent system to test our positive feedback model. Free Energy Rate Density, **Φ**, is measured in MKS units. Those units are different than those used by Chaisson up to a constant, due to the proportionality of the area of the CPUs to their mass. We make the assumption that as a 2D system, the thickness of the silicon wafer is constant, or its change is negligible, and that there are equal amounts of mass per unit area across all generations of CPUs. Therefore, we calculate energy rate density in MKS units $J/s.m^2$, which is analogous up to a constant to the units $J/s.kg$ used previously for FERD and the trends in our data will not change if this constant is used. To calculate the mass per unit area constant, information about the thickness and density of the silicon wafers is necessary. Data only for processors for desktops or laptops were used for consistency, because some of the specialized processors, such as for phones or tablets, perform slower in order to use less energy and fall below this trend line.

Data were collected from Intel Corporation Datasheets [17]. The Instructions Per Second (IPS) for each processor was divided by the Thermal Design Power (TDP) as a measure of the total power consumption by the CPUs at maximum computational speed, for consistency. The result was multiplied by the table value of the Planck's constant, $h = 6.626 \ 10^{-34} \ (Js)$, as the smallest quantum of action, to solve for $\alpha$, as the inverse of the average number of quanta of action per instruction per second [10]. To solve for Q, the TDP was divided by $h$ to find the total number of quanta of action per second. To measure the FERD (**Φ**), we divided the TDP as the maximum rate of energy flowing through the CPUs, by the area (die size) of the CPUs.

## 4 Results

Fig. 2 shows that **Φ** is correlated with the size of the system Q by a power law, which we set to explore with our model. We do not observe large deviations from this power law relation, such as a system with small Q and a large density **Φ** or a system with large Q and small **Φ**. Therefore the total amount of action in a system Q, or its size, is connected to the density of free energy **Φ**, but not just the amount of it. This is a Size-Density rule: size and density in a complex system are proportional. That means mathematically, that **Φ** and α are also in a power law relationship, based on the positive feedback model on Fig. 1 and our previous results for the solutions of the differential equations describing this model [10]. Using this positive feedback cycle, the result is that **Φ** is increasing exponentially in time, which matches with the observations by Chaisson [4]. CPU systems are not observed to have high action efficiency α at low **Φ** and low α at high **Φ**. This proportionality of α with **Φ**, leads to another, Complexity-density rule: complexity (organization, level of development) α is proportional to the time and matter density of free energy, **Φ**, in a system. If one increases or decreases, the other increases or decreases as well.

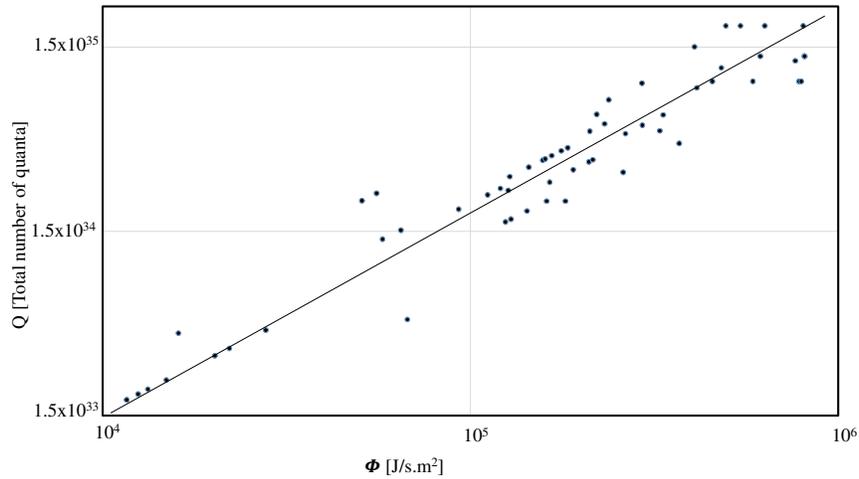

**Fig. 2.** A log-log plot of the total amount of action Q as a function of **Φ**. Data are filled circles and solid line is the fit. The data are from 1982 starting with Intel 286, to 2012, ending with Intel Core i7 3770k. There is a good agreement between the data and a power law fit. The two orders of magnitude change on each axis provide enough data to test the power law relationship between these variables.

# 5 Summary and conclusions

The significance of these calculations is that FERD ($\Phi$) is in a positive feedback loop with organization ($\alpha$) and size (Q) of CPUs and all three reinforce each other, are related with power laws and increase exponentially in time. In order to become better organized, complex systems need larger energy flows to minimize constraints to motion of their elements. Connecting to non-equilibrium thermodynamics, we can say that the further a system is from equilibrium, the more work it can do to minimize constraints to flows, therefore increasing its organization in terms of action efficiency [4,10]. Also, better organization means more action efficient flow channels, therefore higher $\alpha$ provides the necessary efficiency of the flow network to withstand and transmit larger energy flows. As Eric Chaisson points out, at a certain level of structure, the FERD level is an optimum [4]. Too low FERD will slow the system to a stop and too high level will destroy it. That is why we do not find data points above or below the power law trend line. In order to move the optimum level of $\Phi$ higher, the system needs to reorganize and grow in size. This correlations provide observational reason for connecting non-equilibrium thermodynamics with the principle of least action in order to explain progressive increase of organization in complex systems. It agrees with Chaisson's results for Cosmic Evolution, that $\Phi$ grows exponentially in time paralleling the rise in organization [4]. The Least Action Principle (LAP), the Most Action Principle (MAP) and the principle of increase of Free Energy Rate Density (FERD) need to operate together in a positive feedback loop in order to produce an organized complex system.

It remains to be explored if the results are the same for $\Phi$ outside this time interval, for other complex systems and in connection with other characteristics (parameters) of complex systems. If those dependencies hold in other complex systems, they can grow to universal Size-Density and Complexity-Density rules, in addition to the established Size-Complexity rule [10]. We term as "Devology" studies of self-organization in Cosmic Evolution and Development. Our future goal is to study other systems (stellar, physical, chemical, biological, social) for which we can obtain data for $\alpha$, Q and $\Phi$ and to compare with our observations for CPUs. This paper is one step in the further parametrization of the description of the processes of self-organization started earlier [10]. In following research, we plan to add other parameters such as the number of elements, density of elements, number of events and others in the description of the processes of self-organization, and find out if additional regularities exist. As shown in the model if one of the quantities $\alpha$, Q or $\Phi$ increases or decreases, the others increase or decreases predictably and lawfully as well, which is important to take into account in management of complex systems in ecology, engineering, economics, cities and elsewhere in society.

**Acknowledgments.** The authors thank Professor Eric Chaisson, at the Harvard Observatory and Center for Astrophysics (CFA) at Harvard University, for fruitful discussions about Free Energy Rate Density and Cosmic Evolution and Professor Germano Iannacchione, Chair of the Physics department at Worcester Polytechnic Institute about discussions of non-equilibrium systems, as connected to self-organization and FERD. The authors also thank John Smart and Clement Vidal about discussions of the Evolutionary and Developmental processes in the Universe and Assumption College, for financial support and encouragement of this research.